\documentclass[magazine]{IEEEtran}
\IEEEoverridecommandlockouts
\usepackage{cite}
\usepackage[numbers,sort&compress]{natbib}
\usepackage{amsmath,amssymb,amsfonts}
\usepackage{algorithmic}
\usepackage{graphicx}
\usepackage{textcomp}
\usepackage{xcolor}
\def\BibTeX{{\rm B\kern-.05em{\sc i\kern-.025em b}\kern-.08em
		T\kern-.1667em\lower.7ex\hbox{E}\kern-.125emX}}

\usepackage{algorithm} 
\begin{document}
	
	\title{LEO Satellite Access Network (LEO-SAN) Towards 6G: Challenges and Approaches}
	\author{Zhenyu Xiao, \IEEEmembership{Senior Member,~IEEE}, Junyi Yang, Tianqi Mao, \IEEEmembership{Member,~IEEE}, Chong Xu,\\ Rui Zhang, \IEEEmembership{Fellow,~IEEE}, Zhu Han, \IEEEmembership{Fellow,~IEEE}, and Xiang-Gen Xia, \IEEEmembership{Fellow,~IEEE}
		%\thanks{Manuscript received July 7, 2020; revised November 22, 2020 and January 27, 2021; accepted March 1, 2021. This work was supported in part by the National Key R\&D Program of China under Grant 2018YFB1801501 and in part by National Natural Science Foundation of China (Grant No. 61871253). \emph{(Corresponding author: Zhaocheng Wang.)}}
		\thanks{Zhenyu Xiao, Junyi Yang, Tianqi Mao and Chong Xu are with Beihang University; Rui Zhang is with the National University of Singapore; Zhu Han is with University of Houston; Xiang-Gen Xia is with University of Delaware. The corresponding author is Zhenyu Xiao.}}

	\maketitle
	
	\begin{abstract}
		\label{0}
		With the rapid development of satellite communication technologies, the space-based access network has been envisioned as a promising complementary part of the future 6G network. Aside from terrestrial base stations, satellite nodes, especially the low-earth-orbit (LEO) satellites, can also serve as base stations for Internet access, and constitute the LEO-satellite-based access network (LEO-SAN). LEO-SAN is expected to provide seamless massive access and extended coverage with high signal quality. However, its practical implementation still faces significant technical challenges, e.g., high mobility and limited budget for communication payloads of LEO satellite nodes. This paper aims at revealing the main technical issues that have not been fully addressed by the existing LEO-SAN designs, from three aspects namely random access, beam management and Doppler-resistant transmission technologies. More specifically, the critical issues of random access in LEO-SAN are discussed regarding low flexibility, long transmission delay, and inefficient handshakes. Then the beam management for LEO-SAN is investigated in complex propagation environments under the constraints of high mobility and limited payload budget. Furthermore, the influence of Doppler shifts on LEO-SAN is explored. Correspondingly, promising technologies to address these challenges are also discussed, respectively. Finally, the future research directions are envisioned.
		%Their promising solutions and future research directions are also discussed.
		
	\end{abstract}
	%\begin{IEEEkeywords}
	%	LEO-SAN, random access, beam management, doppler-resistant, transmission technologies.
	%\end{IEEEkeywords}
	%	
	
	\section{Introduction}
	\label{1}
	With the commercialization and technology promotion of the fifth-generation (5G) cellular networks, various advanced applications or concepts have been proposed successively, including the industrial Internet of Things (IIoT), holographic communication, digital twins and the meta-universe \cite{c17,c1}. These demands are motivating further technical developments of both industry and academic research, to enter the forthcoming sixth-generation (6G) era \cite{c17,c1}. Compared to the mobile communication technologies up to 5G, the 6G prospects aim to provide ubiquitous coverage and massive user access with higher levels of communication capacity and reliability, which are quite challenging to achieve with the terrestrial cellular networks only. Fortunately, with the rapid development of satellite communication technologies \cite{c1}, the aforementioned requirements for 6G can be satisfied by the incorporation of space-based and ground-based cellular networks \cite{c17,c1,c3,c14,c5}.

	In the multi-tier structure of satellite communication networks, low earth orbit (LEO) satellite constellations play a particularly important role for space-ground interconnection \cite{c3, c14, c5}, and enable the LEO-satellite access network (LEO-SAN) to provide full-coverage broadband services for ground users. Therefore, great attention has been drawn from telecommunication researchers to this paradigm globally. The concept of communicating via LEO satellite constellation can be traced back to the early 1990s. However, due to various issues such as limited technical capability, high cost and insufficient demand, most of LEO constellation programs ended up with bankrupt \cite{c4}. In recent years, the increasing demand for global coverage of broadband Internet access anytime and anywhere has set off a new wave of research and development  on the LEO satellite constellations \cite{c3,c14,c5}. On one hand, in the industry field, there are currently many projects of new-generation large-scale LEO constellations, represented by Iridium II, Starlink, OneWeb, Telesat, and Kuiper, which have made significant progresses, and are beginning for commercial testing \cite{c22,c23}. On the other hand, in the academia, \cite{c3} proposed an enabling network architecture of space-ground integration, under which some key technologies and typical application scenarios were introduced. \cite{c14,c5} took the development trajectory of LEO satellites as the main line, and expounded the research on integration of the LEO-based and terrestrial networks. Furthermore, the 3rd Generation Partnership Project (3GPP) non-terrestrial Network (NTN) Project  is stepping up the standardization of the satellite-ground integrated network. In particular, the REL-17 and subsequent versions require the next period of 5G New Radio (NR) to support LEO satellite communications for extended coverage \cite{c6}. 

	\begin{figure*}[ht]
		\centering
		\includegraphics[width=7in,height=4.25in,clip,keepaspectratio]{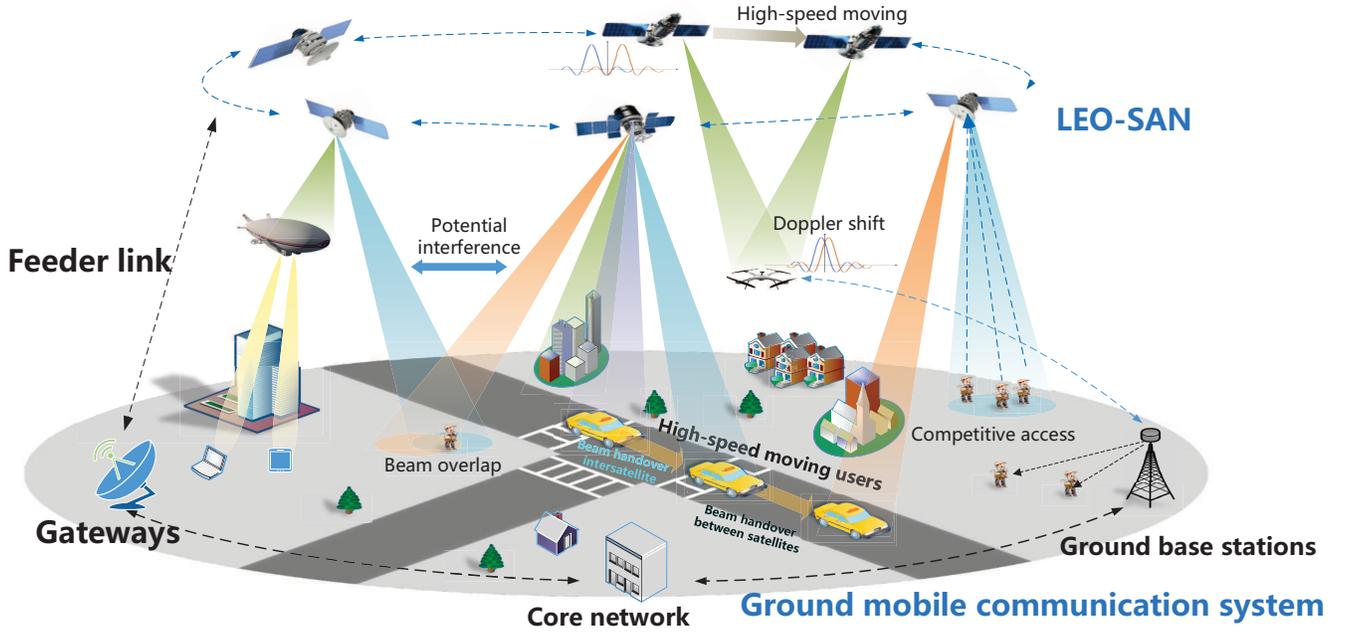}
		\caption{The  typical scenarios and technical composition of LEO-SAN.}
		\label{fig:fig1}
		\vspace*{-2mm}
	\end{figure*}
	
	Fig. \ref{fig:fig1} illustrates the typical scenarios and technical composition of LEO-SAN. The LEO satellites serve as the space base stations (BSs) to complement the terrestrial communication networks, which are capable of attaining full-coverage transmission as well as supporting massive user access with high data rates. Specifically, the user terminals can perform seamless handover between the space and ground BSs in a flexible manner to optimize its quality-of-service (QoS) level, enjoying both the advantages of terrestrial and space-ground data links. Despite its outstanding merits, LEO-SAN still faces several unsolved problems due to its peculiar characteristics of the large space-time scale, high mobility, complex structure and diversified service types. In the aspect of the random access\footnote{Random access in this paper refers to the process from the time when a user sends a random access preamble to the time when a basic signaling connection is established between the user and the BS through information exchange.}, the access architecture is also less flexible. Moreover, the high-speed moving satellite leads to the deterioration of the timeliness of the user's access selection, which may severely affect the access of all kinds of service users. In terms of beam management, the range of beam coverage is inherently contradictory to the access bandwidth. Not only that, limited resources can hardly serve massive users under the existing beam allocation mechanism. As for transmission technology, the high-speed mobility of LEO satellite causes severe Doppler frequency effects, which brings practical challenges to modulation and demodulation.
	
%	\begin{figure*}[ht]
%		\centering
%		\includegraphics[width=7in,height=4.25in,clip,keepaspectratio]{Photo/fig1.eps}
%		\caption{The  typical scenarios and technical composition of LEO-SAN.}
%		\label{fig:fig1}
%	\end{figure*}
%	
	In this work, different from the existing overview papers \cite{c3,c14,c5} that addressed the overall development and application scenarios of LEO-SAN, we will focus on the largely open issues pertaining to random access, beam management and Doppler-resistant transmission technologies for LEO-SAN. More specifically, the main problems of random access in LEO-SAN are discussed including low flexibility, long transmission delay, and inefficient handshakes. Correspondingly, some guidelines for network architecture and algorithm design of random access in LEO-SAN are provided. Then the beam management for LEO-SAN in complicated propagation environments is investigated, where the high mobility and limited payload budget of the LEO satellites are also considered. To guarantee broadband full coverage with reliable data links, advanced beam management solutions are discussed in terms of coverage requirements, beam scheduling, handover management as well as beam resource management. Additionally, we explore the impacts of Doppler shifts on LEO-SAN, where Doppler-resilient signal processing schemes for classical modulation schemes together with novel waveforms against Doppler shifts, are investigated. Finally, some future research directions are discussed.

	\section{Random Access for LEO-SAN}
	\label{2}

	%LEO satellites have limited payload, fast moving speed, limited computing power, long round-trip delay and uneven service distribution. In order to face the above problems and meet the requirements of large bandwidth, low delay and wide coverage of satellite Internet in the future, we will discuss the challenges and possible solutions in the aspects of access architecture, process and the access algorithms of different services. 
	
	In this section, the technical issues in the existing network architecture and random access schemes for LEO-SAN are revealed, including the challenges such as low flexibility, long transmission delay, inefficient handshakes, etc. To tackle these issues, we provide some design guidelines for the network architecture with random access in LEO-SAN. Moreover, we discuss optimized design of random access algorithms for mobile broadband users and burst short packet users, respectively.

	\subsection{Technical Issues in Current Design}
	\label{2A}
	In the existing access network architecture enabled by 3GPP NTN technology, the core network is located on the ground, while the BSs of the access network are all or partially located on the satellite-based platform \cite{c3}. This access architecture has low flexibility and large round-trip latency, which makes it inefficient to support massive users' access. Besides, for mobile broadband service, because of the high speed of satellite movement, the coherent time between the user and BS becomes shorter. Therefore, the timeliness of access user selection in channel state deteriorates, resulting in the decrease of system throughput. In addition, the transmission delay between the satellite and ground user is much longer compared with the ground BS. The existing access mechanism based on multiple handshakes is inefficient under the condition of high latency, and the reliability and continuity of user service cannot be guaranteed. Especially for burst short packet services, re-access after an access failure can cause extra delay and signaling overhead.

	\subsection{Enhanced Network Architecture for LEO-SAN}

	\label{2B}

	\begin{figure}[t]
		\centering
		\includegraphics[width=3.5in,height=4.25in,clip,keepaspectratio]{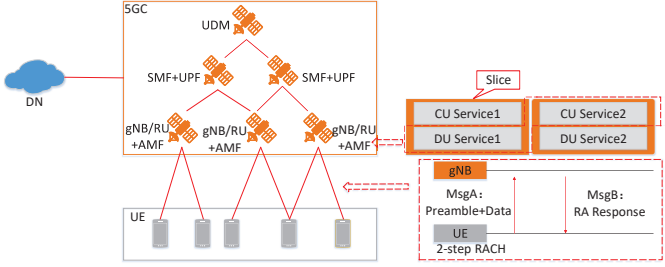}
		\caption{The service-oriented network architecture in LEO-SAN.}
		\label{fig:fig2}
		\vspace{-3mm}
	\end{figure}
	%\begin{figure}[t]
	%	\centering
	%	\includegraphics[width=3in,height=4.25in,clip,keepaspectratio]{Photo/scenario.eps}
	%	\caption{The considered SGIN system.}
	%	\label{fig:scenario}
	%\end{figure}
	%In the existing research on satellite network architecture, all the functions of the core network are located on the ground, and all or part of the access network BSs are located on the satellite platform. Therefore, the data and control signals must go back to the ground for forwarding and processing, which causes the low network flexibility and high end-to-end latency.  With limited load and rapidly changing network topology, it is difficult and challenging to realize low delay, flexible networking and high reliability in spaceborne networks. 
	
	The network architecture of LEO-SAN can be re-designed to address the aforementioned issues. {Fig. \ref{fig:fig2} shows an enhanced service-oriented network architecture with several guidelines provided as follows.} Firstly, the network connection is affected by the dynamic changes of LEO network topology, which thus requires flexible and effective architectural design according to the network environment and users' requirements. The design concept of servitization can be considered in both satellite assess network (SAN) and core network (CN), such that the conventional SAN and CN in 5G can be merged together to facilitate easier network deployment and management. Moreover, network functions or services need to be flexible, distributed, and can be dynamically deployed in multiple nodes of different satellites to achieve efficient coordination of LEO networks. Finally, unified and efficient interface protocols can be designed to connect different network functions under the servitization  framework \cite{c17}. 

	\subsection{Random Access for Mobile Broadband Services}
	\label{2C}
	
	%\begin{figure}[t]
	%	\centering
	%	\includegraphics[width=3in,height=4.25in,clip,keepaspectratio]{Photo/fig2_.eps}
	%	\caption{The random access scenarios of LEO-SAN.}
	%	\label{fig:fig2}
	%\end{figure}
	
	The four-step access method is typically applied for mobile broadband service. Most of the current random access algorithms only focus on the current access moment. However, high-speed mobile satellites and flexible access users result in rapid topological changes. Most of the current access algorithms cannot adapt to the highly dynamic change of the LEO network because these algorithms make decisions only from the current handover moment. {To address this issue, one possible strategy is to utilize the historical state information for future state prediction via efficient learning methods, which can improve the long-term benefits of mobile broadband access. However, user/satellite state information is not only large in scale, but also changes very quickly, leading to lower computational efficiency and worse prediction timeliness. Therefore, how to efficiently  process the massive amount of data {with rapid change} is also essential for performance enhancement of random access. }

	\subsection{Random Access for Burst Short Packet Services}
	\label{2D}
	In LEO-SAN, burst short packet access is one of the most important services of the satellite IoT. Generally, users only send a few bytes of information at a time. The most typical scheme currently studied is random access without authorization, namely the two-step access method \cite{c10}. Compared with the four-step access method, the two-step access method combines the first step and third step to reduce the number of information exchanges and simplify the access process. The preamble and user data have been transmitted together, in the current two-step access method, which helps reduce the signaling complexity and transmission delay. However, the characteristics of channel sparsity and sporadic traffic of massive user devices are ignored. Consequently, a sparse recovery mathematical model can be used to model the problem of joint channel estimation and user detection. {Unfortunately, the problem is usually NP-hard. By introducing certain constraints to the objective function, the problem can be solved by some greedy algorithms and message passing algorithms \cite{c27}.}
	
	%Unfortunately, the problem is usually an NP-hard problem, which may be solved sub-optimally using methods such as maximization-minimization.

	\section{Beam Management in LEO-SAN}
	\label{3}
	
\begin{figure}[t]
		\centering
		\includegraphics[width=3in,height=4.25in,clip,keepaspectratio]{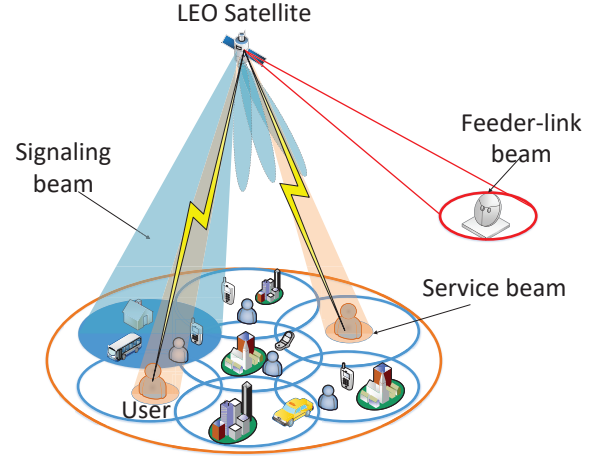}
		\caption{The  typical scenarios beam schedule in LEO-SAN.}
		\label{fig:fig3}
		\vspace{-3mm}
	\end{figure}

	Beam management mainly includes scheduling, handover management and resource allocation of signaling beams, service beams and feeder-link beams, as shown in Fig. \ref{fig:fig3}. In LEO-SAN, the high mobility, complex propagation environment as well as limited budget for communication payloads of the satellite platforms pose great challenges to the beam management operations. This section investigates these technical issues in terms of coverage requirements and beam scheduling, handover management as well as beam resource management.
	
	%Beam management under complex space-ground environment and limited satellite load is the core problem to ensure full-domain broadband coverage. We will discuss coverage requirements and beam schedule, handover management and beam resource management in the following.
	
	\subsection{Coverage Requirement and Beam Scheduling}
	\label{3A}
	
	\subsubsection{Coverage Requirement}
	\label{3A1}
	{In LEO-SAN, there are three important steps for user access: signaling synchronization, service transmission and backhaul. Firstly,  low bandwidth but full coverage are required for signaling synchronization, where the width of signalling beams need to be sufficiently large to cover the whole ground area.  Secondly, the data services require higher bandwidth, and are usually sparsely distributed. Therefore, the service beams only need to point at certain areas of ground users, with reduced beamwidth for higher beamforming gain. Finally, backhaul from LEO satellites to the fixed ground station can be realized by a point-to-point feeder-link beam with even narrower beamwidth and larger communication bandwidth than the counterparts above.}
	
	%Due to the limited payload of LEO satellites, it is of high difficulty for LEO-SAN to provide a sufficient number of beams to cover users in a certain area as traditional geosynchronous earth orbit (GEO) satellite systems. The access control needs to cover all areas to ensure access anywhere, and so the signaling beam in LEO-SAN adopts the combination of beam hopping and beam scanning to achieve the coverage of all areas under LEO satellites. Furthermore, the service distribution is sparse, and the range of beam coverage inherently conflicts contradictory with the access bandwidth. Therefore, the service beam needs to be allocated on demand to improve access bandwidth. As for feeder-link beam, a fixed beam is generally utilized to steadily cover the ground gateway. The following are the specific scheduling techniques for signaling beam, service beam and feeder-link beam

%	The following are the specific scheduling techniques for signaling beam, service beam and feederlink beam.

%	\begin{table}[b]
%		\centering
%		\caption{Requirement comparison of the Three Beams.}
%		\begin{tabular}{|c|c|c|}
%			\hline
%			& Coverage requirment & Beam type  \\
%			\cline{1-3}
%			Signaling beam & Whole domain & Scanning beam \\
%			Service beam & Local area & Spot-beam \\
%			Feeder-link beam & Local area &  Fixed beam  \\
%			
%			\hline
%		\end{tabular}
%		\label{tab:tab1}
%	\end{table}
	
	\subsubsection{Signaling Beam Scheduling}
	\label{3A2}
	{In LEO-SAN, despite the low bandwidth of signaling beams, the link budget is likely to be insufficient due to the limited payload and the high-altitude of LEO satellites. Therefore, the signaling beam may need to cover the ground area in a scanning mode. Each LEO satellite performs a time-division scan of the target area to provide equivalent full coverage. Since each beam scans quickly, the downlink and uplink beams need to be designed independently and spaced at certain intervals to ensure synchronous responses.}
	
	%If the link budget is sufficient, it can directly cover a large area. If the link budget is not sufficient, the signaling beam needs to scan the covered area with a more complex scheduling, i.e., to achieve the same coverage as a single wide beam. In the existing researches, the signal coverage of the beam cell by LEO satellites mainly includes equal area coverage and equal flux coverage. However, due to limited payload capacity and limited beam resources, satellites cannot provide sufficient signaling beams to cover the subsatellite region. Therefore, one possible solution is to design dynamic scheduling strategy of signaling beams to satisfy global users’ access.
	
	%when the satellite load is limited and the beam resources are insufficient, designing a more reasonable signaling beam scanning scheme will be a difficulty and challenge for future research. 
	
	%And random beam is considered as a possible solution to the scanning problem.  
	
	\subsubsection{Service Beam Scheduling}
	\label{3A3}
\begin{figure}[t]
		\centering
		\includegraphics[width=1\linewidth, keepaspectratio]{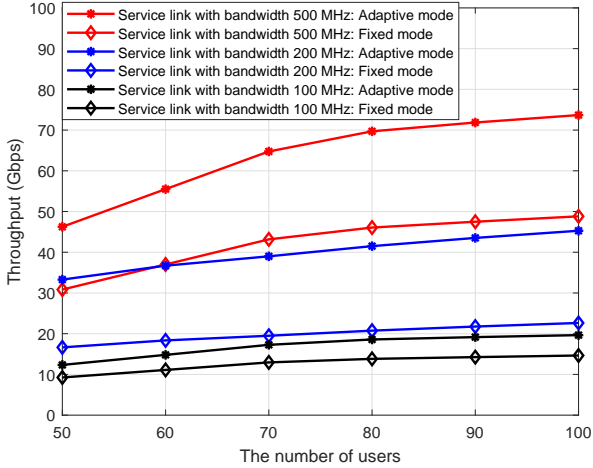}
		\caption{Throughput comparison between the service beams with adaptive-mode and fixed-mode scheduling strategies, under different settings of communication bandwidth.}
		\label{fig:fig4}
		\vspace{-3mm}
	\end{figure}

	{Service beams are mainly utilized to support bandwidth-consuming fixed/mobile services. Since LEO satellites move generally much faster than ground users, the beam scheduling strategies for fixed and mobile services can be similar. Due to payload constraints of the LEO satellites, the key challenge for service beam scheduling is how to utilize a limited number of spot beams to serve a massive amount of users. The possible solutions include beam hopping, effective matching between service beam and signaling beam, and joint scheduling of space, time and frequency resources.} In particular, Fig. \ref{fig:fig4} compares the throughput of service link in a multi-beam LEO-satellite system, using two resource scheduling schemes, i.e., the adaptive mode and fixed mode, under different settings of communication bandwidth.  The satellite in the fixed mode periodically changes beam illumination direction and adopts uniform power allocation strategy. However, the fixed mode does not consider time-varying traffic demand, and fails to fully utilize the space freedom of the beam. In the adaptive mode, we jointly design beam hopping pattern and flexible power allocation according to the distribution of the traffic demand of the ground users, based on matching theory and the successive convex approximation technique. It can be observed that the adaptive beam scheduling mode significantly outperforms the fixed mode under different values of communication bandwidth. This is because the adaptive mode can effectively mitigate the co-channel interference between users, and make full use of the freedom degrees of time, space and transmit power.
	
	%For fixed service, i.e. the user does not move, the satellite directly allocates its service spot beam. For mobile service, satellites need to assign the mobile user a service spot beam that can be tracked. For key service, with the highest priority, the service beam needs to be allocated all the time, even if there is no service transmission. However, in LEO-SAN, a large number of users initiate a large number of service requests, whose density distributions are uneven and quality of service demands are greatly different. Due to limited load capacity and limited bandwidth resources, it is impossible to distribute service beams to all users at the same time. Therefore, it is difficult and challenging for future research to efficiently allocate service spot beams, especially when the specific locationa of users are unknown. Consequently, the efficient matching between service beam and signaling beam is considered as a possible solution to service beam allocation.
	
	\subsubsection{Feeder-Link Beam Scheduling}
	\label{3A4}
	Feeder link is the import and export pipeline of the network, which determines the throughput of the whole network. {The GEO satellites allocate a set of antennas specially for the feeder-link beams.} However, for LEO satellites, the feeder-link beam may share one set of phase-array antenna with the service beam, due to limited budget for communication payload. {In the future, in order to achieve the lightweight and miniaturization of LEO satellites, the integration of service and feeder-link beams is also an important development trend, which also raises the issue of interference between these two types of beams. To reduce their mutual interference, we can optimize the allocation strategy of communication resources for service and feeder-link beams, such as bandwidth, power and center frequency. }

	\subsection{Handover Management}
	\label{3B}
	
\begin{figure}[t]
		\centering
		\includegraphics[width=0.95\linewidth, keepaspectratio]{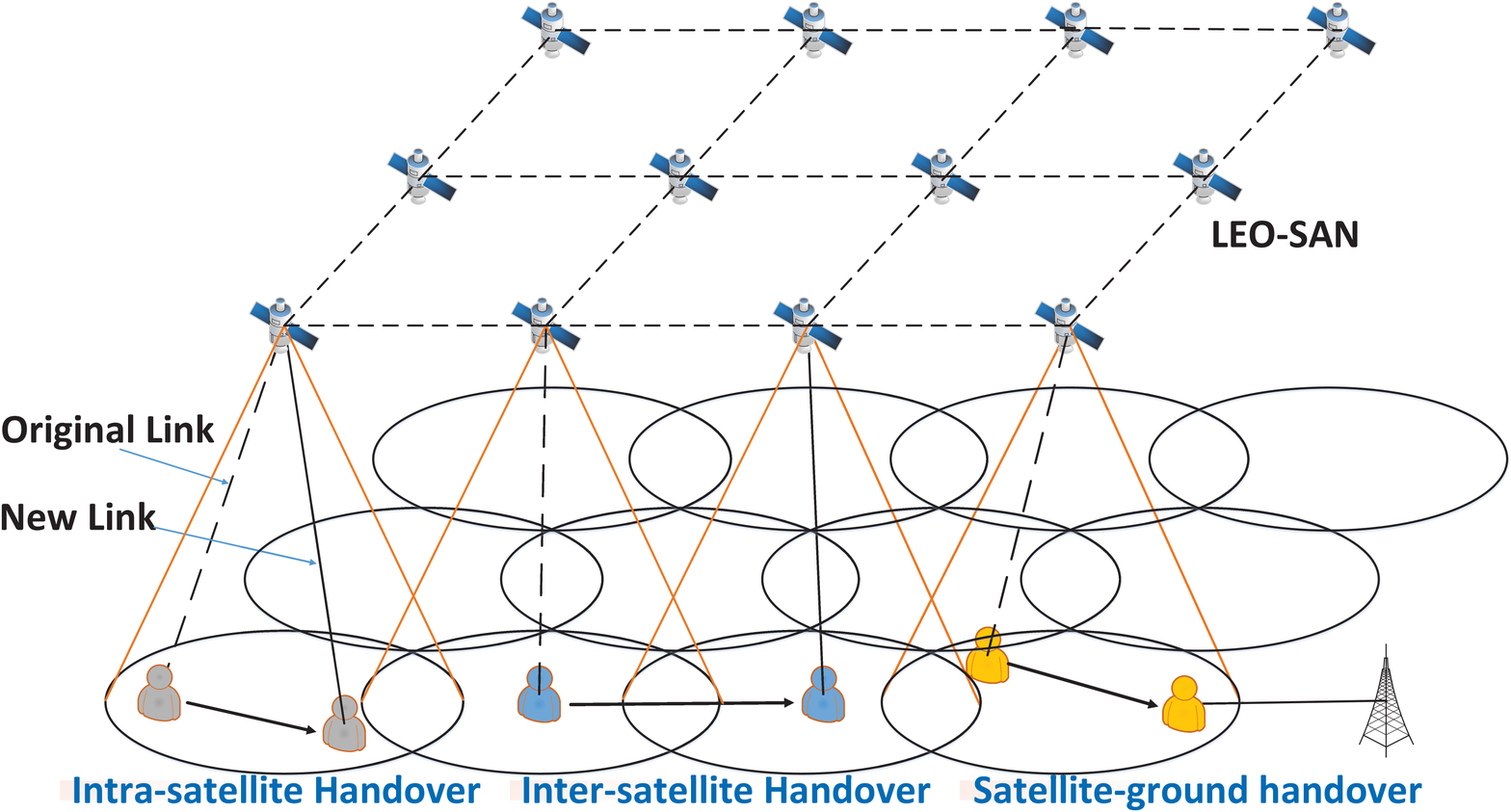}
		\caption{The beam handover scenario of LEO-SAN.}
		\label{fig:fig5}
		\vspace{-3mm}
	\end{figure}

	In LEO-SAN, handover can be classified into intra-satellite beam handover, inter-satellite beam handover and satellite-ground beam handover, as shown in Fig. \ref{fig:fig5}. The high mobility of satellites and uncertainty of massive user movement bring new difficulties and challenges to handover management. It results in frequent beam recasting and handover, which greatly increases signal delay and handover cost, and severely affects user service experience. To address these problems, the handover management strategy can be enhanced from two aspects of handover decision and handover process.
	
	\subsubsection{Handover Decision}
	\label{3B1}
	The handover decision is a critical phase in handover management to determine the user's selection of the next beam. In the literature, the multi-metric handover decision method is widely adopted, by mainly considering receiving signal strength (RSS), service time, the shortest distance, channel resources, available bit rate (ABR), delay, relative speed and network overhead, etc. However, since the traditional method only considers the current state, it is difficult to solve the optimization problem of handover decision that is high-dimensional and highly dynamic, resulting in frequent but unnecessary handovers by users and reduction of the handover success rate. To solve this issue, it is possible to apply more advanced and accurate deep reinforcement learning methods to handover decision. There is no need to know the environment in advance, and the long-term benefits can be maximized by collecting and training experiences and rewards. 
	
	%Our team has made preliminary progress in the research of handover decision algorithm. We apply deep Q-network (DQN) in reinforcement learning to handover decision making. In order to lay a solid theoretical foundation for future research in LEO-SAN, we initially verified the decision algorithm in GEO-SAN scenario including one GEO satellite and multiple ground BSs. As shown in the fig. \ref{fig:fig5}, compared with the traditional methods based on RSS and Q-learning method, DQN method can effectively reduce the frequent handovers of the target user. In the future, we will continue to consider the typical characteristics of high-speed movement of LEO satellites and further apply reinforcement learning methods to LEO-SAN.
	
	%However, the acquisition and efficient processing of high-dimensional and highly dynamic state data are still necessary. 
	
%	\begin{figure}[t]
%		\centering
%		\includegraphics[width=1\linewidth, keepaspectratio]{Photo/fig5.eps}
%		\caption{The handover rate simulaton result.}
%		\label{fig:fig5}
%	\end{figure}
	
	\subsubsection{Handover Process}
	\label{3B2}
	
	The handover process is to define the steps to be performed by the user in the specific handover event. In order to ensure the accuracy of handover information, multi-step information exchange is required. However, different from the access process, the handover process involves the signaling interaction among the original BS, the target BS and the user, as well as the data transmission of the gateway station. 
	{Therefore, for the purpose of improving handover efficiency, on one hand, it is necessary to simplify the handover process as much as possible from the perspective of network architecture. For instance, fusion of CN and RAN architectures can effectively simplify the handover process. On the other hand, the reinforcement learning method considering long-term benefits can be used to make reasonable handover decisions in advance at the user terminals, so that the target BS can configure signaling earlier and realize fast handover.}
	
	%The handover process is more complicated, and it is difficult to improve the handover efficiency by reducing the number of interactive steps. Therefore, how to improve the efficiency and reduce the cost of large-scale user handover events needs careful investigation. For instance, one possible solution is the pre-handover decision method. By introducing long-term benefit methods such as reinforcement learning and utilizing state prediction, users can make reasonable handover decisions in advance, and the target BS can configure signaling in advance to achieve fast handover.
	
	\subsubsection{Group Handover}
	\label{3B3}
	The handover decision algorithm and handover process discussed above are mainly designed for the single-user scenario. Group handover is referred to that a group of users carry out handover together. As one LEO satellite moves out of its original coverage area at high speed, handover will be triggered by numerous users at the same time, which leads to severe handover request signaling storms and a significant increase in the probability of handover process collisions. Therefore, group handover has become a new research focus in LEO-SAN handover management. To avoid the handover request signaling storm on satellites, users can be clustered by location or business, and one representative can be selected as handover-triggering user carrying the information of other users in each group. Besides, in the process of numerous group handover, a high probability of collisions leads to a high handover failure rate. Therefore, the correlations between users, such as their handover sequence, etc., need to be fully exploited, while reinforcement learning method can be used to provide more efficient handover strategies.

    \vspace{-1mm}
	\subsection{Beam Resource Management}
	\label{3C}
	
	%	\begin{figure}[t]
	%		\centering
	%		\includegraphics[width=3in,height=4.25in,clip,keepaspectratio]{Photo/fig6.png}
	%		\caption{Beam resource scheduling in LEO-SAN.}
	%		\label{fig:fig6}
	%	\end{figure}
	%Beam resource management refers to the rational allocation of limited resources to serve more users. In LEO-SAN, power and frequency resources are limited, while the number of beams is much smaller than the number of users. In current, beam-hopping technology is a new satellite beam technology, which can effectively improve the network efficiency by scheduling multi-dimensional resources such as beam, frequency band, time slot and power compared with traditional fixed beam resource allocation strategy. However, the multi-domain resource optimization variable has high dimension, which increases the complexity of constellation resource scheduling. And with the continuous and rapid growth of satellite constellation scale, the potential interference between users increases rapidly. Therefore, in the contradiction between efficient multi-dimensional resource scheduling and limited on-board computing resources under massive heterogeneous services, how to coordinate multi-domain resource scheduling in space, time, frequency and power will be the difficulty and challenge for future research. 
	LEO-SAN usually relies on multi-beam coverage schemes to serve numerous users that may be distant from each other. It requires advanced beam resource management strategies to enhance the network performance. The beam resource management mainly concerns allocation strategies of available beam resources to massive users, which should aim at serving as many users as possible with limited beam resources on LEO satellites. Classical beam resource management schemes employ uniform resource allocation strategy for ground users. However, the uniform allocation method is inefficient under payload constraints and uneven distribution of requirements. Therefore, due to the conflict between efficient multi-dimensional resource scheduling and limited space-borne computing resources, how to coordinate beam resource scheduling in space, time, frequency, power and other domains will be a great challenge for future research. 
	
%	\begin{figure}[t]
%		\centering
%		\includegraphics[width=1\linewidth, keepaspectratio]{Photo/LEO.eps}
%		\caption{The beam handover scenario of LEO-SAN.}
%		\label{fig:fig5}
%	\end{figure}

	\subsubsection{Difficulty of Resource Management and Possible Solutions}
	\label{3C1}
	
	In LEO-SAN, power and frequency resources are limited, while the number of beams is much smaller than the number of users. With the continuous and rapid growth of LEO satellite constellation scale, the variable dimension of multi-domain resource optimization becomes large rapidly, which increases the complexity of constellation resource scheduling. {Beam hopping involves joint optimization of beam pattern design method, power allocation and carrier allocation. We can jointly utilize the convex optimization method, matching theory, machine learning and Lagrange relaxation method for solving beam hopping problems. Compared with the traditional fixed beam resource allocation strategy, it can effectively improve network efficiency and achieve on-demand resource allocation by scheduling multi-dimensional resources such as beam, frequency band, time slot and power.}

	\subsubsection{Interference Management within the Constellation}
	\label{3C2}
	
	LEO-SAN, due to its own unique characteristics and limitations, is bound to receive interference from various interference sources, such as the interference within the small-scale constellation, the interference between the large-scale constellation and the interference in the satellite-ground integrated system. Therefore, in order to support users' various  services reliably, interference management is regarded as one of the key technologies to make full use of the limited frequency and power resources in the integrated network.
	For the interference within a constellation, it can be eliminated by means of pre-coding. As for the interference between constellations, it can be eliminated through proper network planning. Furthermore, one possible solution can be to build a cross-layer system so that GEO satellites can control other LEO systems. Regarding the interference in a satellite-ground integrated system, joint satellite-ground interference management can be carried out when the integrated network operates under spectrum sharing.

	%	Co-channel Interference (CCI) is one of the important factors that restrict the communication capacity of satellite system. When all beams of a multi-beam satellite work together, beams in the same band can interfere with each other, slowing down communication rates. When beam-hopping satellites light up adjacent beams or more beams in the same time slot, there is also mutual interference. Possible solutions are as follows:
	%	\begin{itemize}
	%		\item Multi-beam LEO satellite: In the case of full frequency multiplexing (FFR), CCI between cells can be eliminated by precoding at the transmitter and residual interference can be eliminated by channel equalization at the receiver. In the absence of FFR, interference is eliminated by cell marking and precoding.
	%		\item Beam hopping LEO satellite: Beam hopping is combined with precoding to reduce the impact of CCI. 
	%	\end{itemize}
	%	
	%	\subsubsection{Interference management between constellations}
	%	\label{3C2}
	%	There are many constellation systems in space, not only LEO constellation, but also middle orbit, high orbit and GEO constellation. Interference between different constellations can be eliminated through network scheduling. One possible solution would be to build a cross-layer system so that GEO  satellites could control other LEO systems.
	%	
	%	\subsubsection{Interference by ground network/space-based platform}
	%	\label{3C3}
	%	Joint interference management can be carried out when the integrated network is in spectrum sharing scenario.
	
	\section{Doppler-Resilient Transmission Technology}
	\label{4}
	
	In LEO-SAN, the high mobility of LEO satellites causes severe Doppler shift effects on the signal. For instance, the transmitted signals of LEO-SAN at the s-band experiences Doppler shift up to $\pm48$ kHz, which is much higher than the typical initial UE oscillator inaccuracy (about $20$ kHz) \cite{c14}. The strong Doppler shift can cause undesirable time-selective fast fading, results in performance degradation of channel estimation and signal detection. To this end, the current literature has focused on the estimation and compensation of Doppler shift, mainly based on classical modulation schemes like single-carrier modulation and orthogonal frequency division multiplexing (OFDM) technologies. However, there are still important problems remaining unsolved for effectively dealing with the substantial Doppler frequency shift, such as poor bit-error-rate performance, high inter-carrier interference and low pilot spectrum efficiency. Therefore, some novel Doppler-resilient communication waveforms, including orthogonal time-frequency space (OTFS) modulation \cite{c26} and vector OFDM (VOFDM) \cite{c24}, have been proposed.
	
	%Another way to overcome the Doppler shift effects is new waveform design for Doppler resistance, represented by the orthogonal time-frequency space (OTFS) modulation that, in fact, coincides with  vector OFDM (VOFDM) \cite{c24} at transmission side. 
	
%	\begin{figure}[t]
%		\centering
%		\includegraphics[width=3in,height=4.25in,clip,keepaspectratio]{Photo/fig5.eps}
%		\caption{Doppler-resistant modulation and demodulation technology in LEO-SAN.}
%		\label{fig:fig5}
%	\end{figure}
	%\vspace{-1.5mm}
	\subsection{Doppler-Resilient Transmission Technique for Single-Carrier and OFDM Modulation}
	\label{4A}
	For a single-carrier modulation system, Doppler shift will increase the difficulty of receiver demodulation, and cause performance degradation. For OFDM and other multi-carrier transmission systems, Doppler shift can cause high inter-carrier interference and deteriorate signal detection performance. To solve the above problems, the subcarrier interval can be increased to reduce the interference between subcarriers, or the pilot sequence length can be increased to improve the accuracy of Doppler estimation and compensation. However, the above methods will inevitably lead to the decrease of spectrum efficiency, which is difficult to meet the requirement of massive users for high-rate communication. Moreover, existing Doppler shift estimation and compensation algorithms depend on a high signal-to-noise ratio (SNR) threshold and slow variability of physical channel. Therefore, their performances can be degraded under the high-maneuvering and long-distance scenarios of LEO satellite communications. 
	
	%For the above-mentioned problems, a possible solution can make full use of the inherent pilot information in the data to estimate the Doppler shift through a series of processing, such as rough estimation, matching filtering, detection of implicit pilot, so as to improve the estimation accuracy.
	\vspace{-1mm}
	\subsection{Emerging Modulation Scheme for Doppler Mitigation  }
	\label{4B}
\begin{figure}[t]
			\centering
			\includegraphics[width=1\linewidth,keepaspectratio]{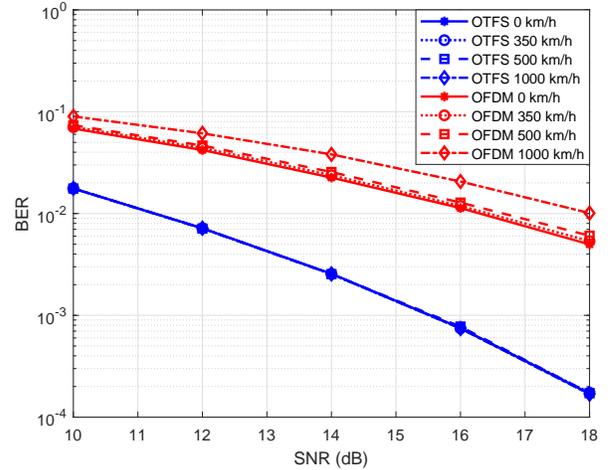}
			\caption{BER comparison of OTFS and OFDM under different levels of Doppler shifts.}
			\label{fig:fig6}
			\vspace{-3mm}
		\end{figure}

	In order to circumvent the challenges to mitigate  Doppler effects using existing modulation methods as illustrated in Section \ref{4A}, a new two-dimensional modulation scheme, namely OTFS, has been proposed \cite{c26}. {Unlike classical time-frequency modulation formats, OTFS carries out modulation in the delay-Doppler domain, which is naturally more suitable for transmission over time-varying wireless propagation channels. This is because, by taking a series of two-dimensional transforms, the time-frequency doubly-dispersive channel is converted to the delay-Doppler domain with almost no channel fading. Fig. \ref{fig:fig6} shows the bit-error-rate (BER) performance comparison between OTFS and classical OFDM under various levels of Doppler shifts. Different relative speeds between the space and ground transceivers, i.e., $0$ km/h, $350$ km/h, $500$km/h and $1,000$km/h, are considered. In simulations, quadrature phase-shift keying (QPSK) is employed for modulation. Besides, the FFT size of OFDM is set as 256, while the size of transmitted symbol matrix in the delay-Doppler domain is set as $256\times14$ for OTFS. It is observed that, on one hand, OFDM suffers from poor communication performance due to strong Doppler shifts. On the other hand, OTFS is capable of achieving significant performance gain over classical OFDM, which validates its superiority in terms of Doppler-shift resistance. Furthermore, almost the same BER performance can be attained by OTFS at different speeds. This further demonstrates its robustness to various levels of Dopper shift effects. However, since the received signals equal the two-dimensional circular convolution between the modulated symbols and the channel coefficients in the delay-Doppler domain, it will be challenging for efficient design of the channel estimators and equalizers. These tasks can be even harder for OTFS systems equipped with large-scale antenna arrays, which is usually practical for LEO-SAN to compensate for the severe path loss.}
	
	%Moreover, it is very promising to combine OTFS and MIMO in the future to combat the strong Doppler effects in LEO-SAN. In order to achieve this combination and make OTFS adapt to the scenario of LEO satellite communications, a possible solution may be to transform the large-scale MIMO problem into a sparse signal estimation problem, which can be solved by the compressed sensing algorithm based on sparse prior conditions.
	
	%However, the existing channel estimation and compensation algorithms of OTFS have not fully explored the time-dependent characteristics and sparsity of the channel, resulting in a large SNR threshold and high computational complexity. In addition, the study on the reception algorithm of MIMO-OTFS is still immature, which makes the existing OTFS technology not suitable for satellite communication scenarios that rely on large-scale antenna arrays. For the above problems, in large-scale MIMO systems, the large-scale MIMO problem can be transformed into sparse signal estimation problem in order to improve transmission performance and channel capacity. Therefore, the compressed sensing algorithm, which relies on sparse prior conditions, is considered as a possible solution. 
	
In contrast, VOFDM \cite{c24} has been proposed as a general format of classical OFDM and single-carrier systems with single transmit antenna, with good robustness to the time-varying channels. Interestingly, the signals in either discrete or continuous format of VOFDM coincide with that in OTFS at the transmitter side. Corresponding to the above example in Fig. \ref{fig:fig6}, the OTFS transmitted sequence is equivalent to the VOFDM counterpart with the vector size of $14$. Next, we provide discussions explaining why OTFS (or VOFDM) is desirable to deal with the Doppler shift effects, from the VOFDM point of view. Explicitly, it has been shown in \cite{c24,c25} that VOFDM can achieve multipath diversity and/or signal space diversity, even with the minimum mean square error (MMSE) linear receiver in a vectorized subchannel \cite{c25}. This is because in VOFDM, at the transmitter side, a vector of information symbols is DFT (or IDFT) transformed implicitly and then, at the receiver side, the information symbols in this vector are demodulated together. More specifically, since in VOFDM, a vectorized channel matrix is pseudo-circulant, it can be diagonalized by DFT/IDFT matrix with a phase shift diagonal matrix (see formula (4.1) in \cite{c24}). Thus, this DFT (or IDFT) processing of a vector of information symbols is similar to the precoding in single antenna systems to attain signal space diversity to combat wireless fading or diagonal space-time block coding in MIMO systems to attain spatial diversity. The resultant diversity gain can effectively mitigate the time-selective channel fading caused by Doppler shifts. For more details, please see \cite{c24,c25}.

	\section{Future Research Direction towards 6G}
	\label{5}
With the continuous commercialization of 5G, 6G research is also in full swing. The LEO-SAN is expected to play an important role for the realization of future 6G wireless networks, which requires extensive further exploration. Therefore, we summarize some possible research directions to provide potential inspirations for the industry and academia.
	
	\subsubsection{Flexible Network Architecture and Virtualized Network Function (VNF) Deployment}
	\label{5A}
	For future development of LEO-SAN, the flexibility of network architecture is required to be further enhanced. In addition to lightweight core network and servitization of access network, technologies such as software-defined network (SDN), network function virtualization (NFV) can also be used to flexibly deploy on-board RAN and CN VNFs, which can support effective load balancing, seamless handover and dynamic resource allocation in LEO-SAN. 
	
	\subsubsection{Efficient Massive Access}
	\label{5B}
	The LEO-SAN is expected to realize massive access for billions of global users in the future 6G networks, which faces a plethora of challenges including the contention for resources request, preamble collision, high signaling overhead, etc. These necessitate advanced technical solutions, e.g., the grant-free random access based on compressive sensing, which are envisioned to break through the user capacity limits of the classical random access counterparts.
	
	%	\subsubsection{Preamble detection}
	%	Narrowband IoT (NB-IoT) is an important application of satellite Internet in the future. In the future, it is inevitable to study a more efficient and accurate preamble detection scheme to deal with massive burst short packet access.
	
%	\subsubsection{Massive Connection}
%	\label{5B}
%	6G is envisioned to provide high-speed Internet access for billions of users around the world. It will become normal for a single satellite to serve a large number of users. Rate splitting multi-access free-grant random access for massive connection in LEO-SAN is thus worth investigating further in the future.
	%has the potential to become a research hotspot in the future.

	\subsubsection{On-demand Beam Management}
	\label{5C}
	%	6G aims to achieve seamless global coverage. Under the condition of limited satellite beam resources, the random beam with space and time scheduling is considered to be an effective solution to the problem of satellite coverage.
	%6G aims to achieve seamless global coverage and has the ability to provide the high quality service which users request. The random beam is a scanning signaling beam and is considered as a solution to global access coverage. In order to enable access users to get accurate service, the matching of random beam and service beam is one of the research focus of future beam management.
	6G aims to achieve seamless global coverage and provide high QoS level for requesting users anytime and anywhere. Therefore, the signaling beams, service beams and feeder-link beams require highly efficient, real-time and dynamic management methods based on specific demands of the user terminals. To this end, the cutting-edge technologies such as artificial intelligence (AI), intelligent optimization, etc., can be introduced to the design of on-demand beam management strategies for LEO-SAN.

	%	\subsubsection{Antenna technology}
	%	Phased array antenna and massive multiple-input multiple-output (MIMO) array will be used to alleviate the shortage of satellite beam.
	
	\subsubsection{Group Handover}
	\label{5D}
	%Under the coverage of high speed change of satellite, massive users clustering to handover is an important way to achieve efficient and seamless handover in the future.
	
	In the future, as more LEO satellites move across each other at high speed, the user handover signaling storm and congestion will  become more severe. To tackle this problem, the combination of group handover and machine learning is regarded as a promising solution to make the handover strategy more adaptable to dynamic scenarios and have more robustness.
	
	%	\subsubsection{Transmission technology}
	%	OTFS, as a new Doppler-resistant transmission technology, combining with MIMO will be more suitable for satellite communication scenarios.
	
	\subsubsection{Intelligent Reflecting Surface (IRS)}
	\label{5E}
	IRS is a large electromagnetic supersurface composed of a large number of low-cost passive reflection elements \cite{c15}, which can reduce the hardware cost of the LEO-satellite-mounted transceiver by replacing the classical phased array antenna. However, the hardware constraints of IRS including the discrete phase shifts and the phase-dependent amplitude coefficient, causes performance degradation of analog beamforming, and poses great difficulties to the usage of quadrature amplitude modulation (QAM), which require detailed investigation.
	
% However, due to the large Doppler effects caused by the high mobility of LEO satellites, in the IRS-aided LEO-SAN, channel estimation and beam tracking will be a challenging problem.
	
	%IRS is a passive array composed of a large number of metamaterial units. It can adjust the electromagnetic wave in wireless environment flexibly with low power consumption and cost, so as to greatly improve the received signal quality \cite{c15}. The combination of IRS and MIMO system will be an emerging solution approach to achieve cost-effective LEO-SAN system in the future. 
	
	\subsubsection{Space-aerial-ground Integrated Network}
	\label{5F}
	LEO-SAN can be incorporated with both the aerial access network and ground-based access network, constituting the promising space-air-ground integrated network (SAGIN), which are expected to reach a new milestone of achieving ubiquitous coverage and superior data throughput. To facilitate its practical implementation, it is necessary to develop advanced strategies of seamless handover among networks, on-demand access, load balancing, connection management, etc.

	%	\begin{itemize} 
	%		\item , more flexible network architecture is needed to be designed. In addition to lightweight core network and servitization of access network, software-defined network (SDN), network function virtualization (NFV) and other technologies can also be used to flexibly deploy on-board network elements to support effective load balancing, seamless movement and dynamic resource allocation in LEO-SAN.
	%		
	%		\item Narrowband IoT (NB-IoT) is an important application of satellite Internet in the future. In the future, it is inevitable to study a more efficient and accurate preamble detection scheme to deal with massive burst short packet access.
	%		
	%		\item 6G will connect hundreds of millions of users around the world. It will become normal for a single satellite to serve a large number of users. Therefore, multi-connection and large-connection access methods will also become a hot research topic in the future.
	%		
	%		\item 6G aims to achieve seamless global coverage. Under the condition of limited satellite beam resources, the random beam with space and time scheduling is considered to be an effective solution to the problem of satellite coverage.
	%		
	%		\item Antenna technology is developing rapidly. In the future, phased array antenna and massive multiple-input multiple-output (MIMO) array will be used to alleviate the shortage of satellite beam.
	%
	%	\end{itemize}
	
	\section{Conclusion}
	\label{6}
	LEO-SAN is expected to become a promising complementary technique to terrestrial cellular networks for future 6G wireless networks, due to its significant  enhancement of capacity and signal coverage. This paper provides an overview of the research development of LEO-SAN. Specifically, the existing problems of LEO-SAN are discussed from the aspects of random access, beam management and Doppler-resilient transmission technologies. These issues are mainly caused by the high mobility of LEO satellites and the long distance of satellite-ground links. This paper also outlines  some possible solutions to tackle these problems and motivate future research. It is hoped that this paper will provide useful inspirations for the development of LEO-SAN technologies.
	
	%	On the basis of the key technologies and research directions proposed in this paper, in order to comply with the development trend and meet the urgent requirements of full coverage, high spectral efficiency, low latency performance and ubiquitous scenarios and applications, researchers need to carry out systematic research on signal and information processing of broadband satellite terminals. In this way, the main performance such as throughput, energy consumption and delay can be improved while ensuring full coverage. 

	%\bibliography{IEEEtran}
	\bibliographystyle{IEEEtran}
	% \bibliographystyle{unsrt}
	% \bibliographystyle{plainnat}
	% \bibliographystyle{elsarticle-num}
	%  \bibliographystyle{elsarticle-num-names}
	% \bibliographystyle{elsarticle-harv}
	% \bibliographystyle{IEEEtran} % use IEEEtran.bst style
	% \bibliography{IEEEabrv,yi_1_1}
	\bibliography{IEEEabrv,ckwx1}

\end{document}